\documentclass[12pt]{article}
\pdfoutput=1
\usepackage{subfigure}
\usepackage{amssymb,amsmath}
\usepackage{graphicx}
\usepackage{color}
\usepackage[colorlinks=true
,urlcolor=blue
,citecolor=blue
,linkcolor=blue
,pagecolor=blue
,linktocpage=true
,pdfproducer=medialab
]{hyperref}
\usepackage[a4paper]{geometry}

\usepackage{placeins}

\makeatletter \renewcommand{\@dotsep}{10000} \makeatother

\def\te{\tilde e}

\def\tu{\tilde u}

\def\tb{\tilde b}

\def\td{\tilde d}
\def\tst{\tilde t}
\def\ttau{\tilde \tau}

\def\tg{\tilde g}

\def\tnu{\tilde\nu}

\def\tq{\tilde q}

\def\mmess{M_{\rm mess}}
\mathchardef\mhyphen="2D
\def\tbta{t\mhyphen b\mhyphen\tau}
\def\btau{b\mhyphen\tau}

\newcommand{\beq}{\begin{equation}}
\newcommand{\eeq}{\end{equation}}
\newcommand{\bea}{\begin{eqnarray}}
\newcommand{\eea}{\end{eqnarray}}

\newcommand{\vev}[1]{\left\langle #1\right\rangle}

\begin{document}

\begin{titlepage}
\pagestyle{empty}

\begin{flushright}
UH-511-1241-15
\end{flushright}

\vspace*{0.2in}
\begin{center}
{\Large\bf    Yukawa Unification and Sparticle Spectroscopy  in Gauge Mediation Models
  }\\
\vspace{1cm}
{\large Ilia Gogoladze$^\ast$\footnote{E-mail: ilia@bartol.udel.edu;\\
\hspace*{0.5cm} On leave of absence from Andronikashvili Institute
of Physics,  Tbilisi, Georgia.},
Azar Mustafayev$^\dagger$\footnote{E-mail: azar@phys.hawaii.edu},
Qaisar Shafi$^\ast$\footnote{E-mail: shafi@bartol.udel.edu} and
Cem Salih $\ddot{\rm U}$n$^{\diamond}\hspace{0.05cm}$\footnote{E-mail: cemsalihun@uludag.edu.tr}}
\vspace{0.5cm}

{\it
$^\ast$Bartol Research Institute, Department of Physics and Astronomy,\\
University of Delaware, Newark, DE 19716, USA \\
$^\dagger$Department of Physics and Astronomy, University of Hawaii, Honolulu, HI 96822, USA \\
$^\diamond$Department of Physics, Uluda\~{g} University, TR16059 Bursa, Turkey
}

\end{center}

\vspace{0.5cm}
\begin{abstract}
\noindent

We explore the implications of $\tbta$ (and $\btau$) Yukawa coupling unification condition
on the fundamental parameter space and sparticle spectroscopy in the minimal gauge mediated
supersymmetry breaking (mGMSB) model.  We find that this scenario prefers values of the CP-odd Higgs mass $m_{A} \gtrsim$ 1~TeV,
with all colored sparticle masses above 3~TeV.
These predictions will be hard to test at LHC13 but they {may} be testable
at HE-LHC 33~TeV or a 100~TeV collider. Both $\tbta$ and $\btau$
Yukawa coupling unifications prefer a relatively light gravitino with mass $\lesssim$ 30~eV,
which makes it a candidate hot dark matter particle. However, it cannot account for more than 15$\%$
of the observed dark matter density.

\end{abstract}

\end{titlepage}


\section{Introduction}
\label{ch:introduction}

The apparent unification of the Standard Model (SM) gauge couplings at the scale
$M_{{\rm GUT}} \simeq 2\times10^{16}$~GeV in the presence of low scale supersymmetry(SUSY)~\cite{book}
is nicely consistent with supersymmetric grand unified theories (GUTs), but it does not significantly
constrain the sparticle spectrum.
On the other hand, imposing $\tbta$ Yukawa coupling unification (YU) condition at $M_{\rm GUT}$~\cite{yukawaUn}
can place significant constraints on the supersymmetric spectrum
in order to fit the top, bottom and tau masses~\cite{Baer:2012cp,Gogoladze:2012ii,Gogoladze:2010fu}.
Most work on $\tbta$
YU condition
has been performed in the framework of gravity mediation SUSY breaking~\cite{Chamseddine:1982jx}.
Some well known choices for GUT-scale boundary conditions on the soft supersymmetry breaking (SSB) terms
yield, with
$\tbta$ YU condition, quite severe constraints on the sparticle
spectrum~\cite{Baer:2012cp,Gogoladze:2012ii,Gogoladze:2010fu,Blazek:2001sb,Auto},
which is further exacerbated after the discovery of the SM-like Higgs boson with mass
$m_h \simeq 125 - 126$~GeV ~\cite{mhatlas,mhcms}.

Models with gauge mediated SUSY breaking (GMSB) provide a compelling
resolution of the SUSY flavor problem, since the SSB terms are generated by the flavor blind gauge
interactions~\cite{Dine:1993yw,Giudice:1998bp,Meade:2008wd}.
In both the minimal~\cite{Dine:1993yw} and general~\cite{Meade:2008wd}  GMSB scenarios, the trilinear SSB A-terms
are relatively small  at the messenger scale, even if an additional sector is added to generate
the $\mu/B\mu$ terms~\cite{Komargodski:2008ax}.
Because of the small A-terms, accommodating the light CP-even Higgs boson mass around 125~GeV requires a stop mass in the
multi-TeV range.  This, in turn, pushes the remaining sparticle mass spectrum to values that are out of reach
of the 14~TeV LHC~\cite{Ajaib:2012vc,mh125spart}. There is hope, however, that the predictions can be tested at 33~TeV  High-Energy LHC (HE-LHC).
This tension between the Higgs mass and the sparticle spectrum can be relaxed if we assume the
existence of low scale vector-like particles that provide significant contribution to the
CP-even Higgs boson mass~\cite{Babu:2008ge,Martin:2012dg}. We will not consider this possibility here.

In this paper we investigate YU in the framework of minimal GMSB (mGMSB) scenario.
Here, the decoupled messenger fields generate SSB terms of particular non-universal pattern. We know from previous studies
in gravity mediation SUSY breaking scenarios that non-universal SSB terms are necessary to achieve successful YU \cite{Gogoladze:2010fu}.
Moreover, below $M_{\rm GUT}$, the mGMSB model contains messenger fields that modify the evolution of the gauge and Yukawa
couplings, thereby affecting YU.
We consider mGMSB in which the messenger fields reside in $5+\bar{5}$ of $SU(5)$, which is the simplest
scenario that preserves unification of the MSSM gauge couplings.
We note that previous studies of mGMSB only considered evolution below the messenger scale, which is usually much lower
than $M_{\rm GUT}$, and therefore could not address the question of YU.

The remainder of this paper is organized as follows. In Section~\ref{sec:gmsb} we briefly outline the mGMSB
framework that we use for our analysis. We describe our scanning procedure along with the experimental constraints
we applied in Section~\ref{sec:constraints}.
In Section~\ref{sec:results} we present our results, focusing in particular on the
 sparticle mass spectrum. The  table in this section presents some benchmark points which summarize
 the prospects for testing these predictions  at the LHC. Our conclusions are presented in Section~\ref{conclusions}.
In Appendix~\ref{sec:appnd},
we briefly discuss the renormalization group equations (RGEs) from the messenger scale to $M_{{\rm GUT}}$.

\section{Minimal Gauge Mediation}
\label{sec:gmsb}

In GMSB models supersymmetry breaking is communicated from a hidden sector to the superpartners of SM particles
via some messenger fields~\cite{Giudice:1998bp}. In the minimal version, the messengers interact with the visible sector only via
SM gauge interactions and to the hidden sector through an arbitrary Yukawa coupling. The {theory is described by} the superpotential
\beq
W = W_{MSSM}+\lambda \hat{S} \Phi \bar{\Phi},
\label{superpot}
\eeq
where $W_{MSSM}$ is the usual superpotential of MSSM, $\hat{S}$ is the hidden sector gauge singlet chiral superfield, and  $\Phi$, $\bar{\Phi}$ are the messenger fields.
In order to preserve perturbative gauge coupling unification, the messenger fields
are usually taken to form complete vector-like multiplets of $SU(5)$. Thus one could have $N_5$ (1 to 4) pairs of $5+\bar{5}$,
or a single pair of $10+\overline{10}$, or the combination $5+\bar{5}+10+\overline{10}$.
For simplicity, we only consider the case with $N_5$ (one to four) pairs of $5+\bar{5}$ vector-like multiplets.

The singlet field $\hat{S}$ develops non-zero VEVs for both its scalar and auxiliary components,
$\vev{\hat{S}}=\vev{S} +\theta^2 \vev{F}$, from the hidden sector dynamics.
This results in masses for the bosonic and fermionic components of the messenger superfields,
\beq
m_b = \mmess\sqrt{1\pm\frac{\Lambda}{\mmess}},\quad m_f = \mmess\, ,
\eeq
where $\mmess=\lambda \vev{S}$ is the messenger scale, and $\Lambda = \vev{F}/\vev{S}$ sets the scale of the SSB terms.
Note that $\vev{F} < \lambda\vev{S}^2$ to avoid tachyonic messengers, and in {many} realistic cases
$\vev{F} \ll \lambda\vev{S}^2$.

Below the messenger scale $\mmess$, the fields $\Phi$ and $\bar{\Phi}$ decouple generating SSB masses for
MSSM fields.
The masses of the MSSM gauginos  are generated at $\mmess$ from 1-loop diagrams with messenger fields. In the
approximation $\vev{F} \ll \lambda\vev{S}^2$,
the gaugino masses are given by
\beq
M_i(\mmess)=\frac{\alpha_i}{4\pi} N_5 \Lambda,
\label{BCgaugino}
\eeq
where $i=1,2,3$ stand for $U(1)_Y$, $SU(2)_L$ and $SU(3)_c$ sectors, respectively.
Since the MSSM scalars do not couple directly to the messengers, their
SSB masses are induced at two loop level:
\beq
m^2_\phi(\mmess)=2 N_5 \Lambda^2 \sum_{i=1}^{3} C_i(\phi) \left( \frac{\alpha_i}{4\pi}\right)^2,
\label{BCscalar}
\eeq
where $C_i(\phi)$ is the appropriate quadratic Casimir associated with the MSSM scalar field $\phi$.

The trilinear A-terms only arise at two-loop level, and thus are vanishing at the messenger scale.
However, they are generated below the messenger scale from the RGE evolution.
{ The bilinear SSB term also vanishes at $\mmess$, although it is often ignored. We do not impose the relation $B=0$,
anticipating that the value needed to achieve the electroweak symmetry breaking (EWSB)
can be explained by some other mechanism operating at the messenger scale~\cite{Komargodski:2008ax}.}

The mGMSB spectrum is therefore completely specified by the following set of parameters:
\beq
\mmess,\ \Lambda,\ N_5,\ \tan\beta,\ sign(\mu),\ c_{grav}\, ,
\label{para}
\eeq
where $\tan\beta$ is the ratio of the two MSSM Higgs doublet VEVs.
The magnitude of $\mu$ is determined by
EWSB condition at the weak
scale. The parameter $c_{grav} (\ge 1$) affects the decay rate of the sparticles into gravitino, but it does not
affect the MSSM spectrum, and we set it equal to unity.

It is worth noting that the gaugino masses in mGMSB are non-universal at the messenger scale (if $\mmess<M_{\rm
GUT}$). However, identical relations at the same scale among the gauginos can be obtained through RGE evolution in the gravity mediated SUSY breaking framework starting with universal
gaugino masses at $M_{\rm GUT}$.
On the other hand, Eq.~(\ref{BCscalar}) implies a universal
SSB mass terms for the up ($H_{u}$) and down ($H_{d}$) Higgs doublets at the messenger scale.
Within the gravity mediated framework this relation can be obtained at the same scale if one imposes
non-universal Higgs masses $m_{H_u}^2< m_{H_d}^2$ at $M_{\rm GUT}$.
Indeed, this non-universality is a necessary condition for compatibility with $\tbta$ YU in
the gravity mediation case with universal gaugino masses at $M_{\rm GUT}$~\cite{Blazek:2001sb,Auto}.
However, the relation $m_{H_u}^2 <  m_{H_d}^2$ is introduced by hand
in gravity mediated SO(10) GUT,
while $\tbta$ YU in mGMSB can be achieved
without such {\it ad hoc} conditions, as we will see later.
From Eq.~(\ref{BCscalar}) the sfermion masses in mGMSB are non-universal at the messenger scale,
but it is not easy to find the same pattern at $\mmess$ within the gravity mediation scenario.
Hence, the sfermion sector could be a good place to see differences between
gravity and gauge mediation SUSY breaking scenarios that are compatible with YU condition at the GUT scale.

\section{Scanning Procedure and Experimental Constraints}
\label{sec:constraints}

For our scan over the fundamental parameter space of mGMSB, we employed ISAJET~{7.84} package~\cite{ISAJET}
that we modified to include the RGE evolution above $\mmess$ in MSSM with $N_5$ pairs of $5+\bar{5}$ messengers.
In this package, the weak-scale values of gauge and third generation Yukawa
couplings are evolved from $M_Z$ to $\mmess$ via the MSSM RGEs in the $\overline{DR}$ regularization scheme.
Above $\mmess$ the messenger fields are present, which changes the beta-functions for the SM gauge and Yukawa couplings.
The modified RGEs between $\mmess$ and $M_{\rm GUT}$ are given in Appendix~\ref{sec:appnd}.
$M_{{\rm GUT}}$ is defined as the scale at which $g_1=g_2$ and is approximately equal to $2\times 10^{16}$~GeV,
the same as in MSSM.
We do not strictly enforce the unification condition $g_3=g_1=g_2$ at $M_{\rm GUT}$, since a few percent deviation
from unification can be accounted by unknown GUT-scale threshold corrections~\cite{Hisano:1992jj}.
For simplicity, we do not include the Dirac neutrino Yukawa coupling
in the RGEs, whose contribution is expected to be small.

The SSB terms are induced at the messenger scale and we set them according to expressions (\ref{BCgaugino}) and
(\ref{BCscalar}). From $\mmess$ the SSB parameters, along with the gauge and Yukawa couplings, are evolved
down to the weak scale $M_Z$.
In the evolution of Yukawa couplings the SUSY threshold
corrections~\cite{Pierce:1996zz} are taken into account at the
common scale $M_{\rm SUSY}= \sqrt{m_{\tst_L}m_{\tst_R}}$,
where $m_{\tst_L}$ and $m_{\tst_R}$
denote the soft masses of the left and right-handed top squarks.
The entire parameter set is iteratively run between $M_Z$ and $M_{\rm
GUT}$ using the full 2-loop RGEs until a stable solution is
obtained. To better account for leading-log corrections, one-loop
step-beta functions are adopted for the gauge and Yukawa couplings, and
the SSB parameters $m_i$ are extracted from RGEs at multiple scales:
at the scale of its own mass, $m_i=m_i(m_i)$, for the unmixed sparticles,
and at the common scale $M_{\rm SUSY}$ for the mixed ones.
The RGE-improved 1-loop effective potential is minimized at $M_{\rm SUSY}$, which effectively
accounts for the leading 2-loop corrections. Full 1-loop radiative
corrections~\cite{Pierce:1996zz} are incorporated for all sparticle masses.


We have performed random scans over the model parameters (\ref{para}) in the following range:
\bea
10^{3}~{\rm GeV}\leq & \Lambda & \leq 10^{7}~{\rm GeV}, \nonumber \\
10^{5} ~{\rm GeV}\leq & \mmess &\leq 10^{16} {\rm GeV}, \nonumber \\
40 \leq & \tan\beta & \leq 60,  \\
N_{5}=1, & \mu < 0. \nonumber
\label{parameterRange}
\eea

Varying $N_{5}$ from 1 to 4, we find that the low scale sparticle spectrum does not change
significantly~\cite{Ajaib:2012vc}. Thus we present results only for $N_{5}=1$ case.
Regarding the MSSM parameter $\mu$, its magnitude but not its sign is determined by the radiative electroweak
symmetry breaking. In our model we set ${\rm sign}(\mu)=-1$.
A negative $\mu$-term together with a positive gaugino mass $M_2$ gives negative
contribution to the muon anomalous magnetic moment calculation~\cite{Moroi}
which would disagree with the measured value for muon $g-2$~\cite{gm2}.
This, however, is not a problem, since the sparticle spectrum is quite heavy in our scenario.
As we will see, the smuons are often heavier than 1~TeV. Hence, the SUSY contribution to  muon $g-2$ is not significant,
which also is the case in $\tbta$ YU scenarios~\cite{Baer:2001kn,Gogoladze:2010fu}.
On the other hand, a negative sign of $\mu$ with positive signs for all gauginos helps to realize YU.
The reason for this is that in order to implement YU, we require significant SUSY threshold corrections
to the bottom quark Yukawa coupling.
One of the dominant finite corrections is proportional to $\mu M_3$~\cite{Hall}  and
it helps to realize Yukawa unification if this combination has negative sign~\cite{Gogoladze:2010fu}.
Finally, we employ the current central value for the top mass, $m_{t}=173.3$~GeV.
Our results are not too sensitive to one or two sigma variation
of $m_{t}$~\cite{Gogoladze:2011db}.

In scanning the parameter space, we employ the Metropolis-Hastings
algorithm as described in Ref.~\cite{Belanger:2009ti}. The data points collected all satisfy
the requirement of radiative electroweak symmetry breaking (REWSB). After collecting the data, we impose
the mass bounds on  the particles~\cite{pdg} and use the IsaTools package~\cite{isatools,Baer:2001kn}
to implement the various phenomenological constraints. We successively apply mass bounds
including the Higgs~\cite{mhatlas,mhcms} and gluino masses~\cite{gluinoLHC},
and the constraints from the rare decay processes $B_s \rightarrow \mu^+ \mu^-$~\cite{BsMuMu},
$b \rightarrow s \gamma$~\cite{Amhis:2012bh} and $B_u\rightarrow\tau \nu_{\tau}$~\cite{Asner:2010qj}.
The constraints are summarized below in Table~\ref{table1}.
Notice that we used wider range for the Higgs boson mass, since there is an approximate error of around 2~GeV in the estimate of the mass
that largely arises from theoretical uncertainties in the calculation of the minimum of the
scalar potential, and to a lesser extent from experimental uncertainties in the values of $m_t$ and $\alpha_s$.

\begin{table}[h]\centering
\begin{tabular}{rlc}
$   123\, {\rm GeV} \leq m_h \leq127$ \,{\rm GeV} &
\\
$ m_{\tilde{g}} \geq1.5$ \,{\rm TeV} & \\
$0.8\times 10^{-9} \leq{\rm BR}(B_s \rightarrow \mu^+ \mu^-)
  \leq 6.2 \times10^{-9} \;(2\sigma)$ &
\\
$2.99 \times 10^{-4} \leq
  {\rm BR}(b \rightarrow s \gamma)
  \leq 3.87 \times 10^{-4} \; (2\sigma)$ &
\\
$0.15 \leq \frac{
 {\rm BR}(B_u\rightarrow\tau \nu_{\tau})_{\rm MSSM}}
 {{\rm BR}(B_u\rightarrow \tau \nu_{\tau})_{\rm SM}}
        \leq 2.41 \; (3\sigma)$ &
\end{tabular}
\caption{Phenomenological constraints implemented in our study.}
\label{table1}
\end{table}

Following ref.~\cite{Belanger:2009ti} we define the parameters $R_{tb\tau}$ and $R_{b\tau}$ which quantify $\tbta$ YU
and $\btau$ YU respectively:
\begin{equation}
R_{tb\tau} = \frac{{\rm Max}(y_{t},y_{b},y_{\tau})}{{\rm Min}(y_{t},y_{b},y_{\tau})},
\hspace{0.5cm}
R_{b\tau} = \frac{{\rm Max}(y_{b},y_{\tau})}{{\rm Min}(y_{b},y_{\tau})}
\end{equation}
{Thus, $R$ is a useful indicator for Yukawa unification with $R\lesssim 1.1$,
for instance, corresponding to Yukawa unification to within 10\%, while
$R=1.0$ denotes `perfect' Yukawa unification.}

\section{Results}
\label{sec:results}

\begin{figure}[h!]
\centering
\includegraphics[scale=1]{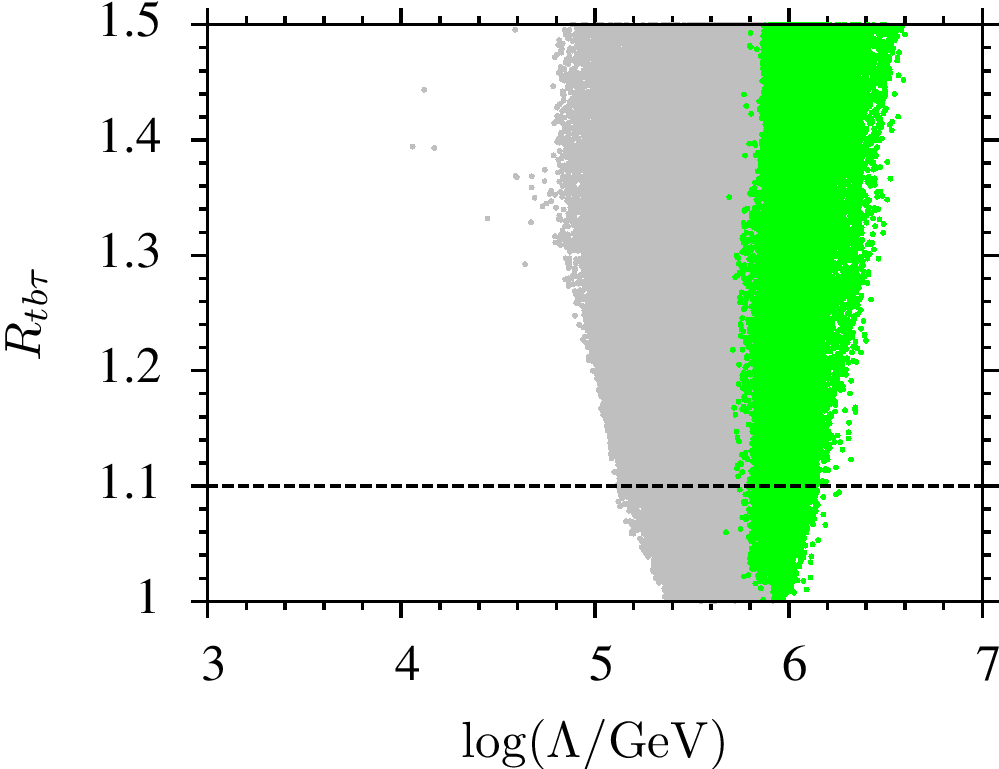}\hfill
\includegraphics[scale=1]{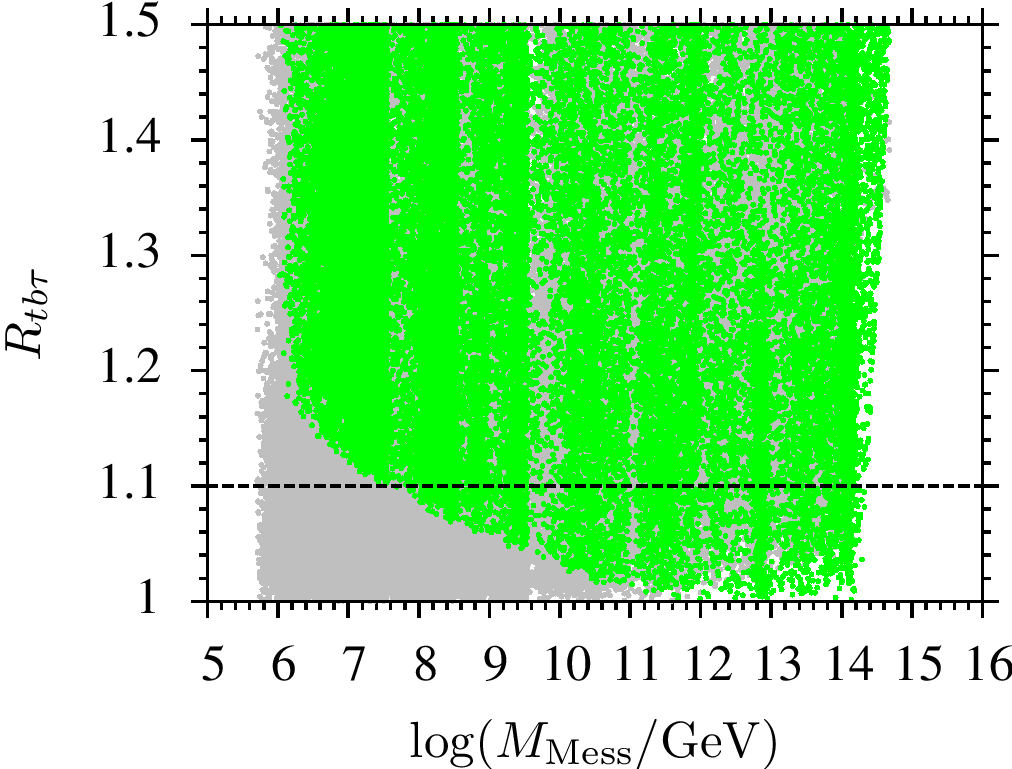}\hfill
\includegraphics[scale=1]{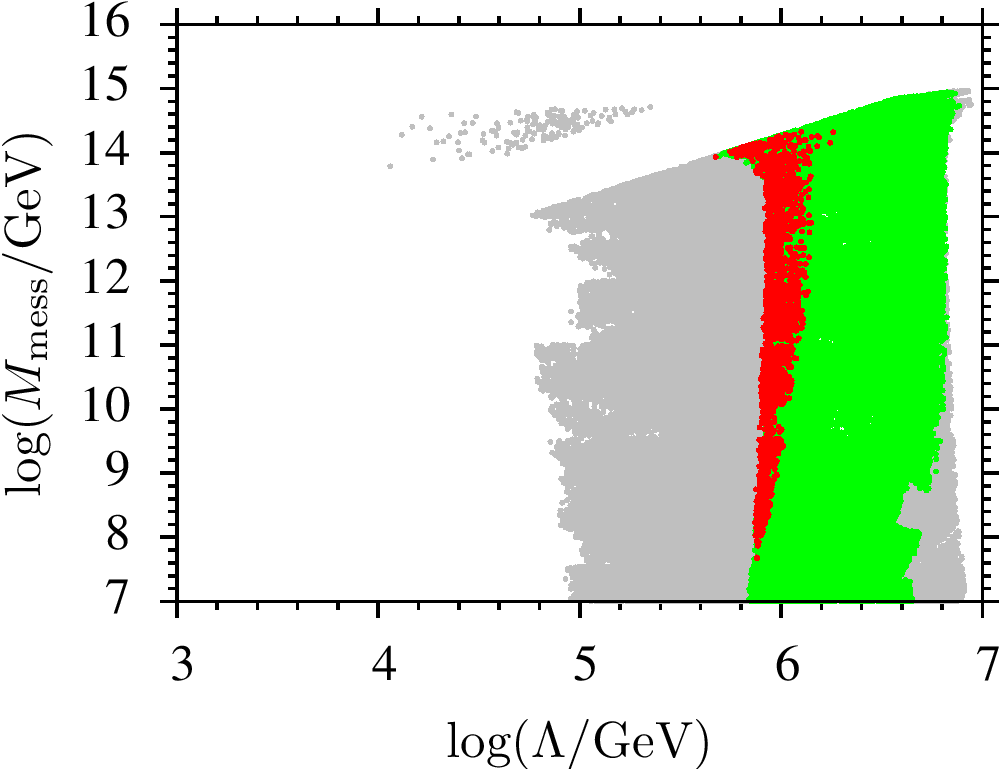}\hfill
\includegraphics[scale=1]{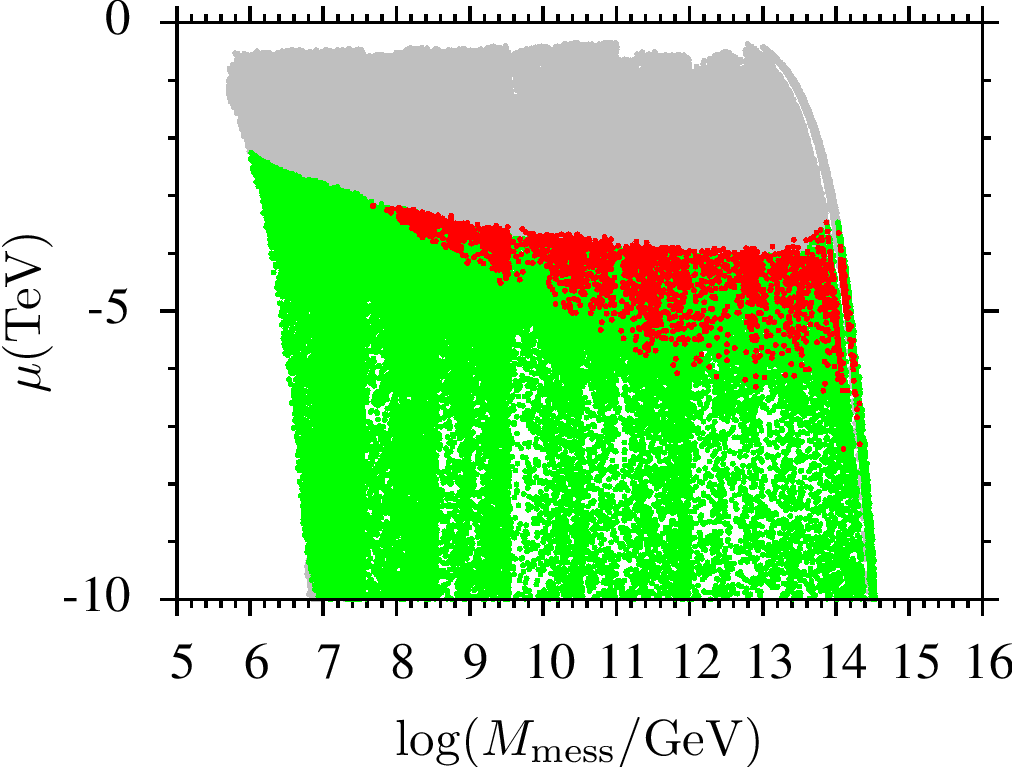}
\caption{Plots in $R_{tb\tau}-\Lambda$, $R_{tb\tau}-\mmess$, $\mmess-\Lambda$ and $\mu - \mmess$
planes. All points are consistent with REWSB. Green points also satisfy mass bounds and B-physics constraints.
Red points are a subset of green and they are compatible with t-b-$\tau$ YU.
In the top panels regions below the horizontal line are compatible with YU such that $R_{tb\tau}\leq 1.1$}.
\label{figure1}
\end{figure}

\begin{figure}[h!]
\includegraphics[scale=1]{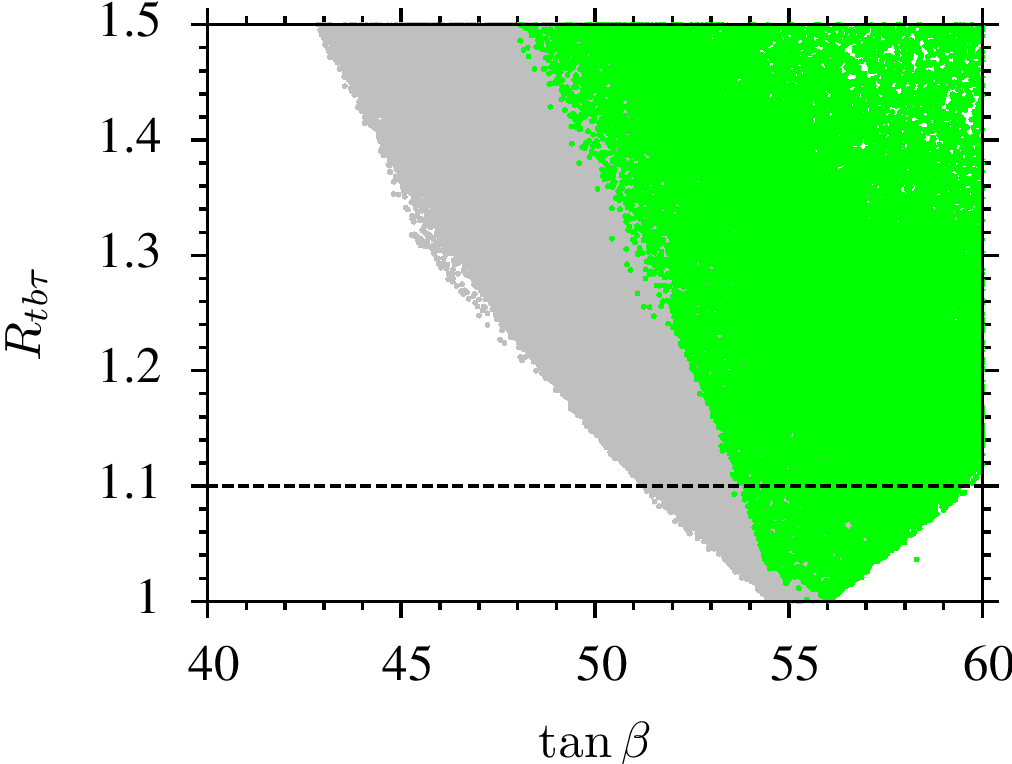}\hfill
\includegraphics[scale=1]{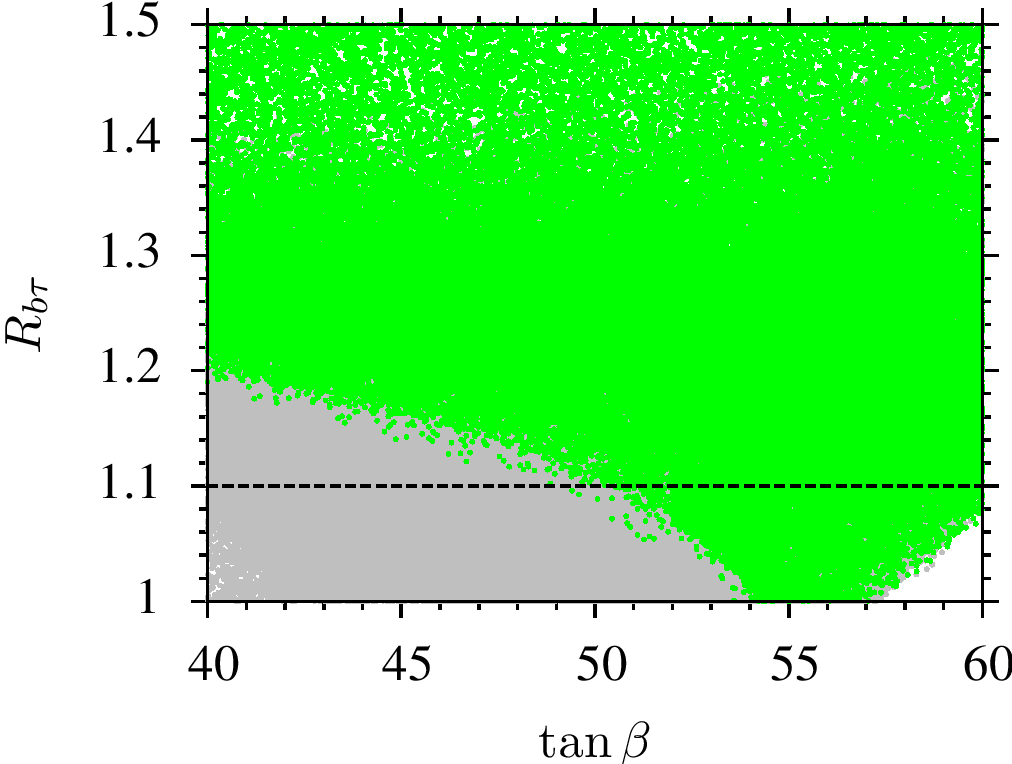}

\caption{Plots in $R_{tb\tau}-\tan\beta$ and $R_{b\tau}-\tan\beta$ planes.
The color coding is the same as in Figure~\ref{figure1}.}
\label{figure2}
\end{figure}

\begin{figure}[h!]
\includegraphics[scale=1]{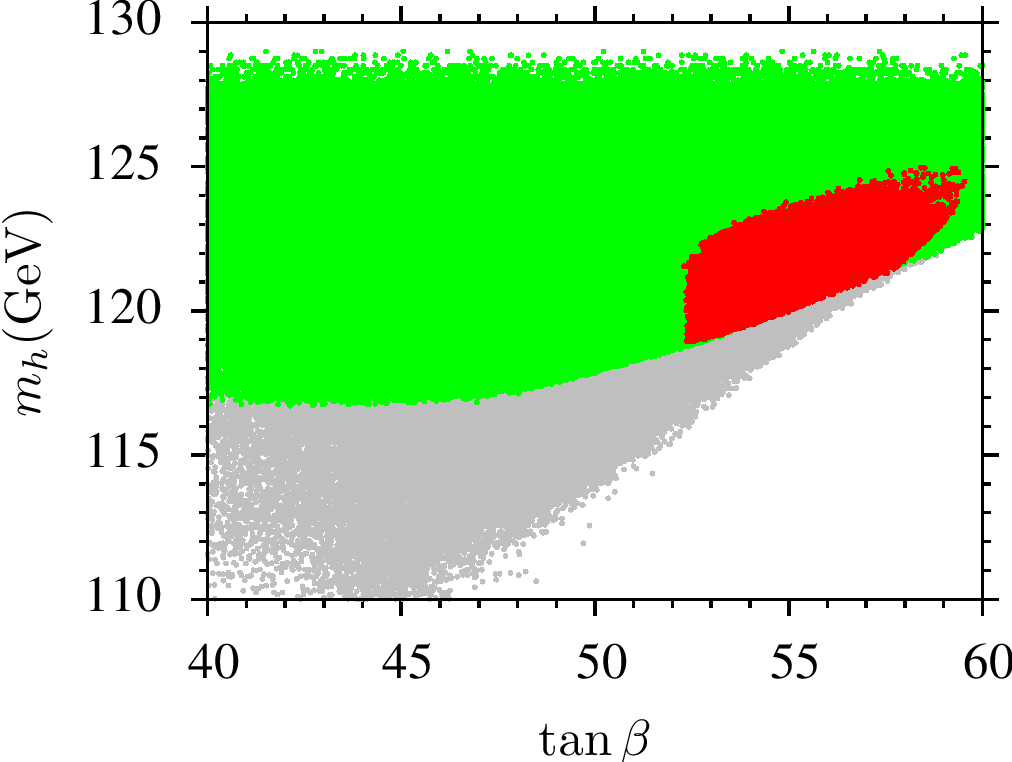}\hfill
\includegraphics[scale=1]{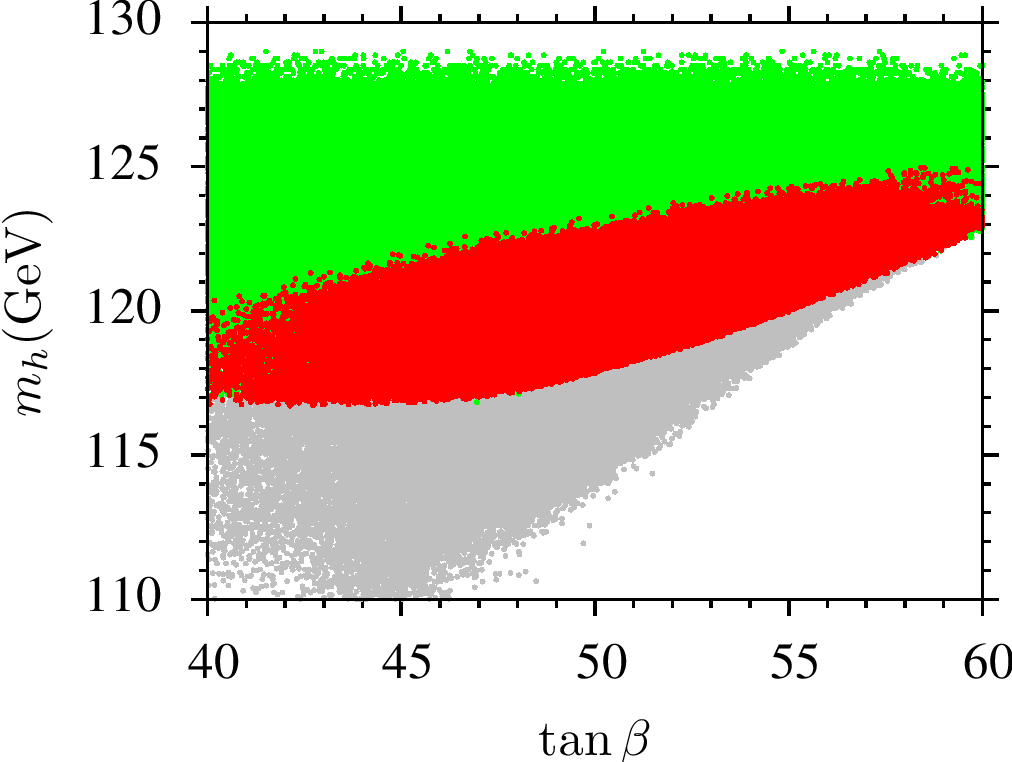}
\caption{Plots in $m_h -\tan\beta$  plane for $\tbta$ YU (left panel) and $\btau$ YU (right panel).
The color coding is the same as in Figure~\ref{figure1} except the Higgs mass bound is not applied in these panels.}
\label{figure3}
\end{figure}

\begin{figure}[h!]
\includegraphics[scale=1]{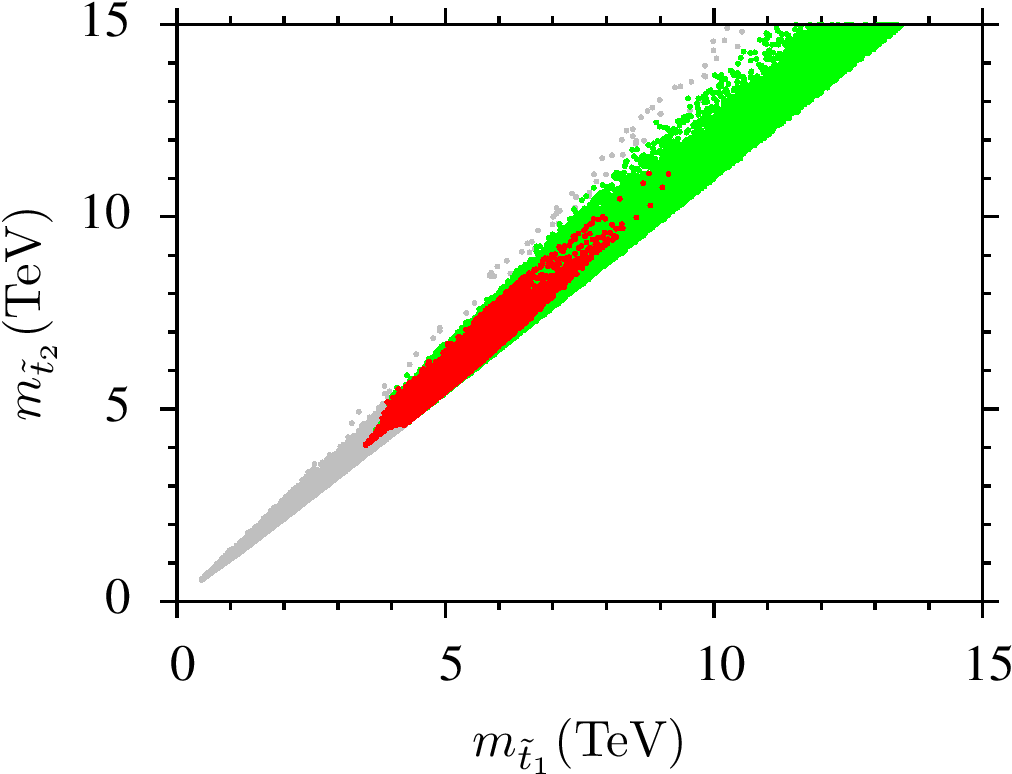}\hfill
\includegraphics[scale=1]{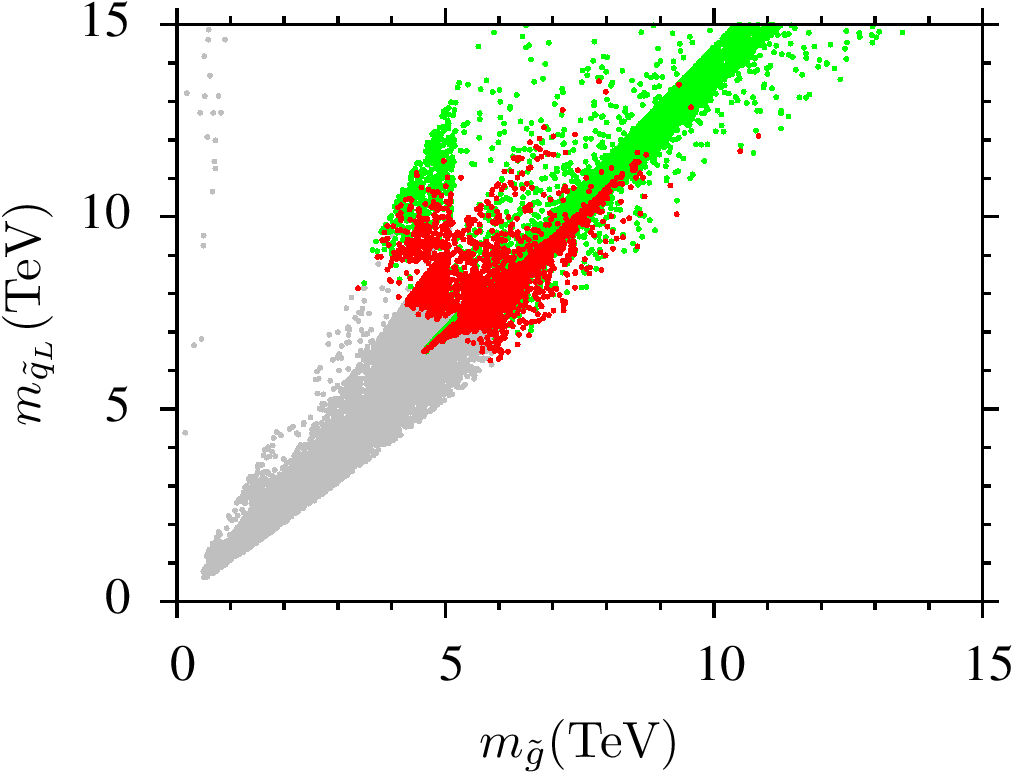}\hfill
\includegraphics[scale=1]{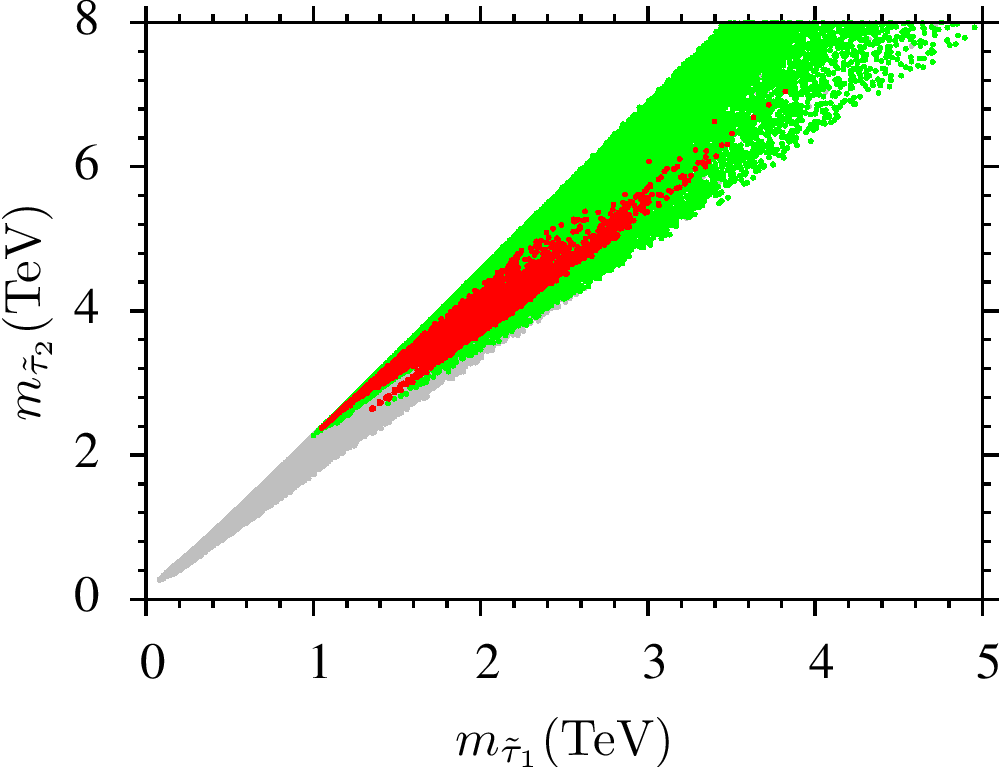}\hfill
\includegraphics[scale=1]{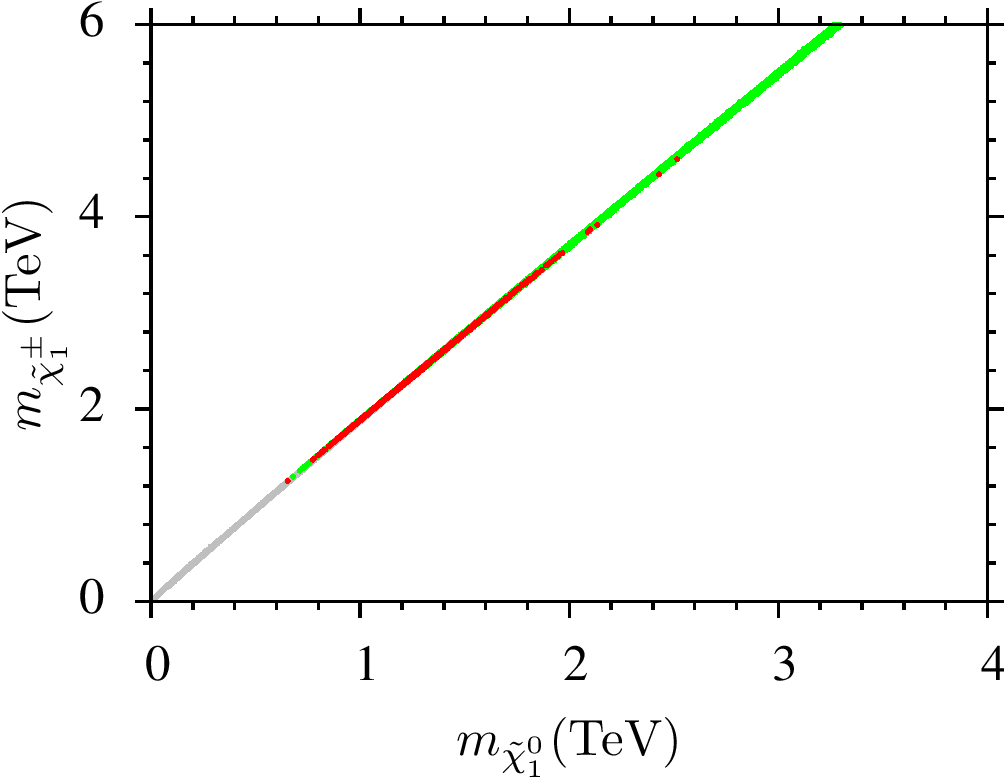}
\caption{Plots in $m_{\tst_{1}}-m_{\tst_{2}}$, $m_{\tu_{L}}-m_{\tg}$,
$m_{\ttau_{1}}-m_{\ttau_{2}}$ and $m_{\tilde{\chi}^{\pm}_{1}}-m_{\tilde{\chi}^{0}_{1}}$ planes.
The color coding is the same as in Figure~\ref{figure1}.}
\label{figure4}
\end{figure}

\begin{figure}[h!]\hspace{0.1cm}
\includegraphics[scale=1.05]{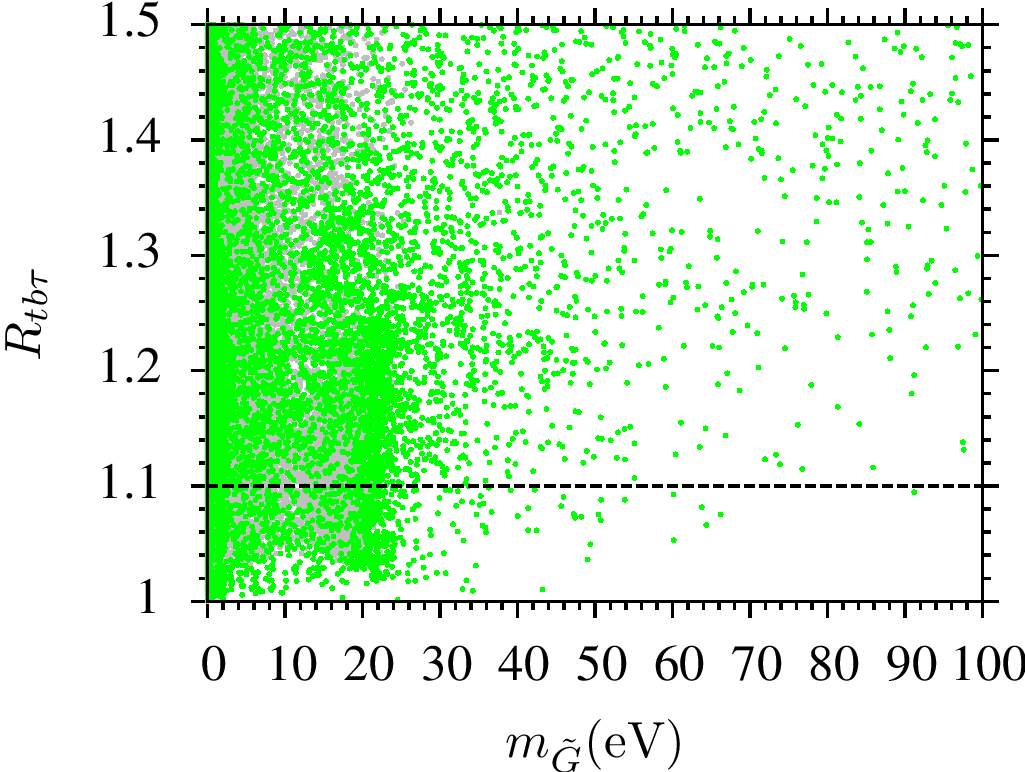}\hfill
\includegraphics[scale=1.05]{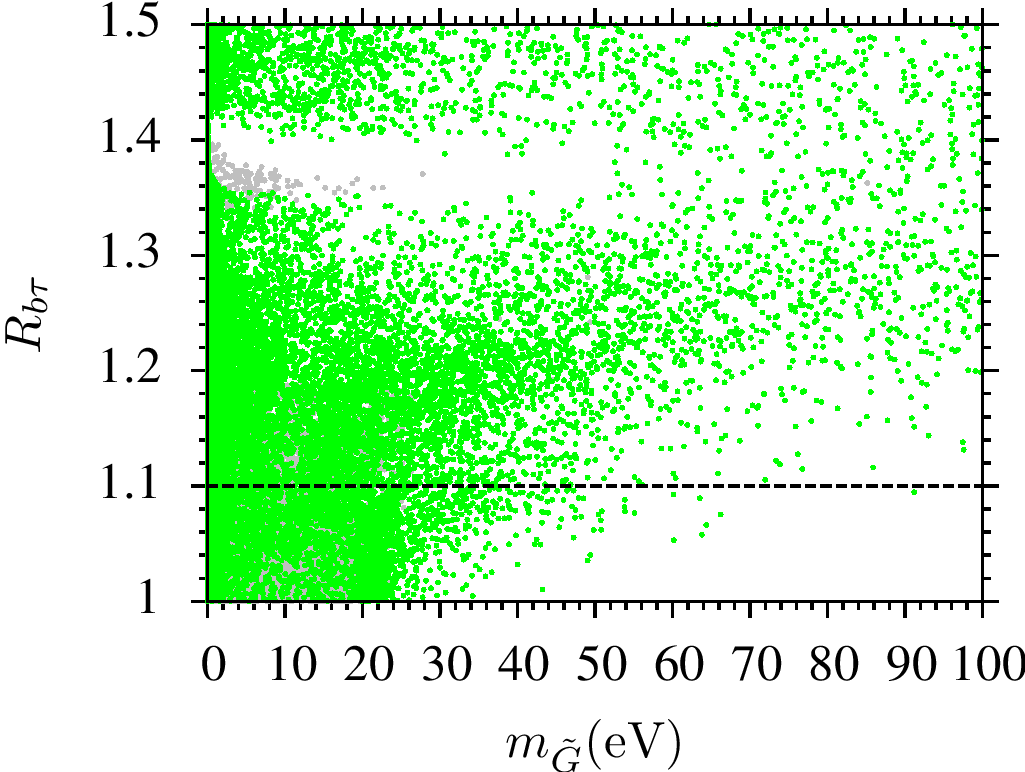}
\caption{Plots in $\mmess-m_{\tilde{G}}$, $\Lambda-m_{\tilde{G}}$, $R_{tb\tau}-m_{\tilde{G}}$ and
$R_{b\tau}-m_{\tilde{G}}$ planes. The color coding is the same as in Figure~\ref{figure1}.}
\label{figure5}
\end{figure}

\begin{table}[t!]
\centering
\scalebox{0.89}{
\begin{tabular}{|c|ccccc|}
\hline
                 & Point 1 & Point 2 & Point 3 & Point 4 & Point 5 \\
\hline
\hline
$\Lambda$       & $0.94\times 10^{6}$  & $1.51\times 10^{6}$ & $7.6\times 10^{5}$ & $4.7\times 10^{5}$ & $1.16\times 10^{6}$ \\
$\mmess$ & $1.5\times 10^{14}$ & $1.3\times 10^{14}$ & $1.2\times 10^{8}$ & $8.6\times 10^{13}$ & $1.3\times10^{15}$\\
$\tan\beta$     & 56.1 & 59.3 & 58.7 & 54.8 & 52.7\\
\hline
$\mu$          & -4906 & -6727 & -2995 & -4268 & -3984\\
$A_{t}$        & -3564 & -5479 & -1933 & -1834 & -4513\\
$A_{b}$        & -3600 & -5540 & -1946 & -1891 & -4868\\
$A_{\tau}$     & -268.1 & -462.9 & -123.2 & -140.4 & -535.1\\
\hline
$m_h$           & 124 & 125 & 123 & 123 & 124\\
$m_H$           & 1148  & 1640 & 929 & 1665 & 2378\\
$m_A$           & 1141 & 1629 & 923 & 1654 & 2363\\
$m_{H^{\pm}}$   & 1153 & 1643 & 934.2 & 1668 & 2380\\

\hline
$m_{\tilde{\chi}^0_{1,2}}$
                 & 1305, 2438  &  2098, 3868  &  1045, 1971 &  654, 1249 & 1602, 2960\\

$m_{\tilde{\chi}^0_{3,4}}$
                 & 4902, 4903 & 6725, 6725  & 3001, 3002 & 4190, 4190 & 3993, 3995\\

$m_{\tilde{\chi}^{\pm}_{1,2}}$
                & 2440, 4902 & 3871, 6724 & 1972, 3002 & 1251, 4154 & 2961, 3996\\

$m_{\tg}$  & 6113 & 9350 & 4995 & 3372 & 7178\\
\hline $m_{ \tu_{L,R}}$
                 & 9545, 8590  &  13434, 12152 & 6903, 6469 & 8140, 7206 & 7818, 7255\\
$m_{\tst_{1,2}}$
                 &6325, 7784 & 9152, 11112 & 5569, 6186 &  5011, 6515 & 5752, 6745\\
\hline $m_{ \td_{L,R}}$
                 & 9545, 8317 & 13434, 11796  & 6903, 6412 & 8140, 6937 & 7819, 7124\\
$m_{\tb_{1,2}}$
                 & 6280, 7743  & 9134, 11053  & 5617, 6148 & 5054, 6482 & 6022, 6709\\
\hline
$m_{\tnu_{e,\mu}}$
                 & 5181 & 7122 & 2727 & 4659 & 3560\\
$m_{\tnu_{\tau}}$
                 & 4843 &  6625 & 2659 & 4358 & 3372\\
\hline
$m_{ \te_{L,R}}$
                & 5186, 3698 & 7129, 5044  & 2735, 1478 & 4663, 3324 & 3567, 2378\\
$m_{\ttau_{1,2}}$
                & 2600, 4845 & 3397, 6628 & 1193, 2667 & 2340, 4357 & 1748, 3381\\
\hline
$R_{tb\tau}$ &1.00 & 1.09  & 1.09 & 1.06 & 1.23\\
$R_{b\tau}$ &1.00 & 1.08  & 1.06 & 1.02 & 1.07\\
\hline
\end{tabular}}
\caption{Benchmark points for exemplifying our results. All masses and scales are given in GeV units.
Point 1 depicts a solution with perfect YU. Point 2 shows a solution with 125~GeV mass for h consistent with YU. Point 3 displays
a solution with essentially the lowest stau mass consistent with YU.
Similarly, point 4 shows essentially the lightest mass for the lightest neutralino.
This point also demonstrates the lightest gluino and stop masses obtained from our scans.
Point 5 displays a solution which is compatible with $\btau$ YU,
but not with $\tbta$ YU.}
\label{benchgmsb}
\end{table}

Figure~\ref{figure1} displays
the regions in the parameter space that are compatible with $\tbta$ YU with plots in
$R_{tb\tau}-\Lambda$, $R_{tb\tau}-\mmess$, $\mmess-\Lambda$ and
$\mu - \mmess$ planes. All points are consistent with REWSB. Green points satisfy the mass bounds
and B-physics constraints. Red points are a subset of green ones and they are compatible
with $\tbta$ YU. In the top panels the  dashed line indicates
the region consistent with YU with $R_{tb\tau}\leq 1.1$. The $R_{tb\tau}-\Lambda$ plane shows that the parameter
$\Lambda$ in mGMSB is constrained by
$10^{5}\,\rm{GeV} \lesssim \Lambda \lesssim 10^{6}$~GeV for $\tbta$ YU with $R_{tb\tau} \leq 1.1$.
As seen from the $R_{tb\tau}-\mmess$ plane, YU imposes a far less stringent constraint on $\mmess$,
namely $10^{8}\,\rm{GeV} \lesssim \mmess \lesssim 10^{14}$~GeV.
We also present, for additional clarity, the results in $\mmess-\Lambda$ plane.
The MSSM parameter $\mu$ lies in the range $-8\,\rm{TeV} \lesssim \mu \lesssim -3$~TeV for compatibility with YU.
We see that there are plenty of solutions consistent with the mass bounds and constraints
from rare B-decays. However, only the thin red stripe is compatible with YU.

The strong impact of $\tbta$ YU on the fundamental parameter space is somewhat relaxed if only $\btau$ YU is imposed.
Figure~\ref{figure2}  shows the differences
between $\tbta$ and $\btau$ YU cases with plots in $R_{tb\tau}-\tan\beta$ and
$R_{b\tau}-\tan\beta$ planes. The color coding is the same as in Figure~\ref{figure1}.
One sees that, $\tbta$ YU only allows a very narrow range
for $\tan\beta$, namely $54 \lesssim \tan\beta \lesssim 60$, and perfect
$\tbta$ YU consistent with the experimental constraints can be realized when $\tan\beta \approx 56$. On the other hand,
the $R_{b\tau}-\tan\beta$ plane shows that the range for
$\tan\beta$ is slightly enlarged, $50 \lesssim \tan\beta \lesssim 60$,
with $\btau$ YU, and perfect $\btau$ YU is obtained
for $54 \lesssim \tan\beta \lesssim 57$.

A similar comparison based on the range for the SM-like Higgs boson mass
is given in Figure~\ref{figure3} with plots in $m_h -\tan\beta$  planes for $\tbta$
(left panel) and $\btau$ (right panel) YU cases.
The color coding is the same as in Figure~\ref{figure1}, except that the Higgs mass bound is not applied
in these panels. The left panel shows that demanding $10\%$ or better YU,
the lightest CP-even Higgs boson mass can be $ m_{h} \approx 119-125$~GeV for $\tbta$ case,
while it can be as light as 117~GeV in the case of $\btau$ YU, as shown in the right panel.
The panels of Figure \ref{figure3} highlight how stringent the Higgs boson mass constraint is
on the fundamental parameter space of mGMSB compatible with YU.
The left panel shows that the 125~GeV Higgs boson mass and $\tbta$ YU more or less equivalently constrain the parameter space.
On the other hand, as seen from the right panel, $\btau$ YU can be realized for $\tan\beta \gtrsim 40$,
while the mass bound on the Higgs boson excludes the region with $\tan\beta \lesssim 50$.

Figure~\ref{figure4} displays the mass spectrum of supersymmetric particles in
$m_{\tst_2}-m_{\tst_1}$, $m_{\tu_L}-m_{\tg}$,
$m_{\ttau_2}-m_{\ttau_1}$ and
$m_{\tilde{\chi}^{\pm}_{1}}-m_{\tilde{\chi}^{0}_{1}}$ planes,
with the color coding the same as in Figure~\ref{figure1}. As seen from the top panels,
the colored sparticles are quite heavy. Thus $\tbta$ YU requires
$m_{\tilde{t}_1} \gtrsim 4$~TeV.  Since the SSB trilinear scalar interaction term
$A_t$ is relatively small~\cite{Ajaib:2012vc}, the heavier stop is expected to be about
the same mass ($\gtrsim 4$ TeV) as the lighter stop.
Similarly, the squarks of the first two families and the gluino have masses in the
few TeV range. The $m_{\tq_L}-m_{\tg}$ plane shows that $m_{\tg} \gtrsim 3$~TeV
and $m_{\tq_L} \gtrsim 6$~TeV for consistency with $\tbta$ YU as well as
the experimental constraints.

The bottom panels of Figure~\ref{figure4} display the mass spectra
for the lightest sparticles which are found to be stau, chargino and neutralino.
According to the $m_{\ttau_2}-m_{\ttau_1}$ panel, $\ttau_1$ can be as
light as 1~TeV, while $m_{\ttau_2} \gtrsim 2$~TeV.
Similarly, the $m_{\tilde{\chi}^{\pm}_{1}}-m_{\tilde{\chi}^{0}_{1}}$ plane displays the masses
of the lightest chargino and neutralino with the latter $\gtrsim 600$~GeV, and the former $\gtrsim 1200$~GeV.

Finally, Figure~\ref{figure5}  displays the gravitino mass,
$m_{\tilde{G}}$, in $R_{tb\tau}-m_{\tilde{G}}$ and $R_{b\tau}-m_{\tilde{G}}$ planes with the color coding as in Figure~\ref{figure1}. As is well-known,
the gravitino is usually the LSP in the mGMSB framework, and its mass can be as heavy as 10~TeV.
A light gravitino is a plausible dark matter candidate and it can also manifest itself through
missing energy in colliders~\cite{Feng:2010ij}.

In the standard scenarios, the relic density bound ($\Omega h^2 \simeq 0.11$~\cite{Komatsu:2008hk})
is satisfied with a gravitino mass $\sim$ 200~eV~\cite{Feng:2010ij},
which makes it a hot dark matter candidate.
The hot component of dark matter, however, cannot be more than $15\%$ which, in turn, implies
that the gravitino mass should be less than about $30$~eV~\cite{Feng:2010ij}. This indeed is predicted
from Figure~\ref{figure5}.
The panels of Figure~\ref{figure5} represent the impact of YU on the gravitino mass.
Both $\tbta$ and $\btau$ YU do not allow a gravitino that is too heavy,
preferring instead a gravitino lighter than 100~eV or so. The $R_{tb\tau}-m_{\tilde{G}}$ plane
shows that $\tbta$ YU is mostly realized with $m_{\tilde{G}} \lesssim 30 $~eV,
with essentially perfect YU possible for nearly massless gravitinos.
The bound on the gravitino mass is not significantly relaxed for $\btau$ YU.
However, essentially perfect YU is found for $m_{\tilde{G}} \lesssim 50$~eV.
In order to have a complete dark matter scenario one could invoke axions as cold dark matter.

In Table~\ref{benchgmsb} we present five benchmark points
that exemplify our results. The points are chosen to be consistent with the
experimental constraints and YU. The masses are given in GeV. Point 1 depicts a solution with perfect YU.
Point 2 shows the heaviest mass in our scan for the lightest CP-even Higgs boson, while
point 3 displays the lightest stau mass that we found, consistent with YU.
Similarly, point 4 represents the lightest neutralino mass foundin our scan.
This point also shows the lightest gluino and stop masses that we obtained.
Point 5 displays a solution which is compatible with $\btau$ YU, but not with $\tbta$ YU.

\section{Conclusion}
\label{conclusions}

{ Models based on the gauge mediated SUSY breaking are highly motivated, as they provide an atractive solution
to the SUSY flavor problem.}
We have explored the implications of $\tbta$ and $\btau$ YU condition on the sparticle spectroscopy in the framework
of mGMSB { where the messengers reside in $5+\bar{5}$ multiplets of $SU(5)$ and only interact with visible sector via gauge interactions.}
We find that YU leads to a heavy sparticle spectrum with the lowest mass sparticles being the lightest neutralino
($m_{\tilde{\chi}_{1}^{0}}\gtrsim 600$~GeV), chargino ($m_{\tilde{\chi}_{1}^{\pm}}\gtrsim 1200$~GeV)
and stau ($m_{\ttau} \gtrsim 1000$~GeV). In addition, YU prefers values of the CP-odd Higgs boson mass
$m_{A}$ to be greater than 1~TeV and all colored sparticle masses above 3~TeV.
Such heavy sparticles can escape detection at the 13-14~TeV LHC,
but should be accessible at HE-LHC 33~TeV or the proposed 100~TeV collider.
{ We find that YU requirement in the  mGMSB framework leads to  a spectrum of gauginos, Higgses and staus, that is very similar to the one  in  gravity mediated $SO(10)$ GUT, although for different reasons.
The details of the rest of sfermion spectrum are different, but they are at about the same heavy scale.
However, YU in mGMSB framework does not require any {\it ad hoc} assumptions, such as as SSB non-universalities that are necessary in
gravity mediated GUTs.}
It is also interesting to note that $\tbta$ and $\btau$ YU favor a gravitino mass less than 30~eV,
which makes it a hot dark matter candidate.
In this case, the gravitino cannot comprise more than 15\% of the dark matter density,
and one may invoke axions in order to have a complete dark matter scenario.

\section*{Acknowledgments}
We thank David Shih for useful discussions.
AM would like to thank FTPI at the University of Minnesota for hospitality during the final stages of this
project.
This work is supported in part by DOE Grants DE-FG02-91ER40626 (IG and QS) and DE-SC0010504 (AM).
This work used the Extreme Science and Engineering Discovery Environment (XSEDE),
which is supported by the National Science Foundation Grant Number OCI-1053575.
Part of the numerical calculations reported in this paper were performed at the National Academic Network and
Information Center (ULAKBIM) of Turkey Scientific and Technological Research Institution (TUBITAK),
High Performance and Grid Computing Center (TRUBA Resources).
IG acknowledges support from the Rustaveli National Science Foundation under grant No.~31/98.

\appendix
\section{RGEs for MSSM and $5+\bar{5}$}
\label{sec:appnd}

Above the messenger scale $\mmess$ the MSSM is supplemented with vector-like messenger fields and is
described by the superpotential in Eq.~(\ref{superpot}).
Since the theory is supersymmetric above $\mmess$, there are no SSB
terms in the lagrangian. The $N_{5}$ messenger pairs $5+\bar{5}$ of SU(5) decompose under the SM
gauge group as
\beq
\Phi+\bar{\Phi}=L_M\left(1,2,\frac{1}{2}\right)+\bar{L}_M\left(1,2,-\frac{1}{2}\right)
               +D_M\left(\bar{3},1,\frac{1}{3}\right)+\bar{D}_M\left(3,1,-\frac{1}{3}\right),
\eeq
where numbers in parentheses represent transformation properties under $SU(3)_c$, $SU(2)_L$ and $U(1)_Y$ respectively.

At the 1-loop level the MSSM gauge couplings contain additional term $g^3_a N_5$, where $a=1,2,3$ correspond to the gauge groups
$U(1)_Y$, $SU(2)_L$ and $SU(3)_c$, respectively. At the 2-loop level additional contributions appear from
terms where the Dynkin index is summed over all chiral multiplets. Hence the gauge coupling RGEs have the form
\beq
\frac{d g_a}{dt}=\frac{1}{16\pi^2}\left(\beta^{(1)}_a+g^3_a N_5\right)
        +\frac{1}{(16\pi^2)^2}\left(\beta^{(2)}_a+g^3_a \sum_{b=1}^3 2N_5  B^{(2)}_{ab}(5) g^2_b\right).
\eeq
Here, $\beta^{(1,2)}_a$ are one- and two-loop MSSM beta functions that can be found in Ref.~\cite{rges},
and
\beq
B^{(2)}_{ab}(5)=
\begin{pmatrix}
7/30 & 9/10 & 16/15 \\
3/10 & 7/2 & 0\\
2/15 & 0 & 17/3
\end{pmatrix}
\eeq
is the contribution from a single 5-plet. Note that $5$ and $\bar{5}$ give the same contribution to RGEs.

The anomalous dimensions for MSSM fields are modified only at the 2-loop level, since messengers only couple to
MSSM fields via gauge interactions. The RGEs for the MSSM Yukawa couplings are
\bea
\frac{d y_t}{dt}&= & \frac{1}{16\pi^2}\beta^{(1)}_{t}
                 +\frac{1}{(16\pi^2)^2}\left(\beta^{(2)}_t
		     +y_t N_5\left(\frac{8}{3}g_3^4+\frac{3}{2}g_2^4+\frac{13}{30}g_1^4 \right) \right),\\
\frac{d y_b}{dt}&= & \frac{1}{16\pi^2}\beta^{(1)}_{b}
                 +\frac{1}{(16\pi^2)^2}\left(\beta^{(2)}_b
		     +y_b N_5\left(\frac{8}{3}g_3^4+\frac{3}{2}g_2^4+\frac{7}{30}g_1^4 \right)\right),\\
\frac{d y_{\tau}}{dt}&= & \frac{1}{16\pi^2}\beta^{(1)}_{\tau}
                   +\frac{1}{(16\pi^2)^2}\left(\beta^{(2)}_\tau
		     +y_\tau N_5\left(\frac{3}{2}g_2^4+\frac{9}{10}g_1^4 \right)\right),
\eea
where $\beta^{(1,2)}_i$ are one- and two-loop MSSM beta functions~\cite{rges}.



\begin{thebibliography}{99}


\bibitem{book}
H.~Baer and X.~Tata, {\it Weak Scale Supersymmetry: From Superfields to Scattering Events},
(Cambridge University Press, 2006);
  S.~P.~Martin,
  Adv.\ Ser.\ Direct.\ High Energy Phys.\  {\bf 21}, 1 (2010)
  [hep-ph/9709356];
 M.~Drees, R.~Godbole and P.~Roy,
 {\it Theory and phenomenology of sparticles: An account of four-dimensional N=1
  supersymmetry in high energy physics}, (World Scientific,2004).

\bibitem{yukawaUn}
B. Ananthanarayan, G. Lazarides and Q. Shafi, Phys.\ Rev.\ D {\bf 44},1613 (1991) and
Phys.\ Lett.\ B {\bf 300}, 24 (1993)5;
Q.~Shafi and B.~Ananthanarayan, Trieste HEP Cosmol.1991:233-244.

\bibitem{Baer:2012cp}
  H.~Baer, S.~Raza and Q.~Shafi,
  Phys.\ Lett.\ B {\bf 712}, 250 (2012);
  A.~Anandakrishnan, B.~C.~Bryant, S.~Raby and A.~Wingerter,
  Phys.\ Rev.\ D {\bf 88}, 075002 (2013).

\bibitem{Gogoladze:2012ii}
  I.~Gogoladze, R.~Khalid and Q.~Shafi,
  Phys.\ Rev.\ D {\bf 79}, 115004 (2009);
  S.~Antusch and M.~Spinrath,
  Phys.\ Rev.\ D {\bf 79}, 095004 (2009);
  I.~Gogoladze, Q.~Shafi and C.~S.~Un,
  JHEP {\bf 1207}, 055 (2012);
  I.~Gogoladze, Q.~Shafi and C.~S.~Un,
  JHEP {\bf 1208}, 028 (2012);
  M.~Badziak,
  Mod.\ Phys.\ Lett.\ A {\bf 27}, 1230020 (2012);
  A.~S.~Joshipura and K.~M.~Patel,
  Phys.\ Rev.\ D {\bf 86}, 035019 (2012);
  A.~Anandakrishnan and S.~Raby,
  Phys.\ Rev.\ Lett.\  {\bf 111}, 211801 (2013);
  M.~Badziak, M.~Olechowski and S.~Pokorski,
  JHEP {\bf 10}, 088 (2013);
  M.~A.~Ajaib, I.~Gogoladze and Q.~Shafi,
  Phys.\ Rev.\ D {\bf 88}, 095019 (2013);
  S.~Antusch, S.~F.~King and M.~Spinrath,
  Phys.\ Rev.\ D {\bf 89}, 055027 (2014);
  M.~A.~Ajaib, I.~Gogoladze, Q.~Shafi and C.~S.~Un,
  JHEP {\bf 1405}, 079 (2014);
  T.~Li, D.~V.~Nanopoulos, S.~Raza and X.~C.~Wang,
  JHEP {\bf 1408}, 128 (2014).

\bibitem{Gogoladze:2010fu}
  I.~Gogoladze, R.~Khalid, S.~Raza and Q.~Shafi,
  JHEP {\bf 1012}, 055 (2010);
 M.~Adeel Ajaib, I.~Gogoladze, Q.~Shafi and C.~S.~Un,
  JHEP {\bf 1307}, 139 (2013).

\bibitem{Chamseddine:1982jx}
 A.~Chamseddine, R.~Arnowitt and P.~Nath, Phys.\ Rev.\ Lett.\ {\bf 49} (1982) 970;
R.~Barbieri, S.~Ferrara and C.~Savoy, Phys.\ Lett.\ {\bf B119} (1982) 343;
N.~Ohta, Prog.\ Theor.\ Phys.\ {\bf 70} (1983) 542;
L.~J.~Hall, J.~D.~Lykken and S.~Weinberg, Phys.\ Rev.\ {\bf D27} (1983) 2359.

\bibitem{Blazek:2001sb}
  T.~Blazek, R.~Dermisek and S.~Raby,
  Phys.\ Rev.\ Lett.\  {\bf 88}, 111804 (2002).

\bibitem{Auto}
  H.~Baer and J.~Ferrandis,
  Phys.\ Rev.\ Lett.\  {\bf 87}, 211803 (2001);
  D.~Auto, H.~Baer, C.~Balazs, A.~Belyaev, J.~Ferrandis and X.~Tata,
  JHEP {\bf 0306}, 023 (2003).

\bibitem{mhatlas}
  G.~Aad {\it et al.}  [ATLAS Collaboration],
  Phys.\ Lett.\ B {\bf 716}, 1 (2012).

\bibitem{mhcms}
  S.~Chatrchyan {\it et al.}  [CMS Collaboration],
  Phys.\ Lett.\ B {\bf 716}, 30 (2012).

\bibitem{Dine:1993yw}
  M.~Dine and A.~E.~Nelson,
  Phys.\ Rev.\ D {\bf 48}, 1277 (1993);
  M.~Dine, A.~E.~Nelson and Y.~Shirman,
  Phys.\ Rev.\ D {\bf 51}, 1362 (1995);
  M.~Dine, A.~E.~Nelson, Y.~Nir and Y.~Shirman,
  Phys.\ Rev.\ D {\bf 53}, 2658 (1996).

\bibitem{Giudice:1998bp}
  For a review, see G.~F.~Giudice and R.~Rattazzi,
  Phys.\ Rept.\  {\bf 322}, 419 (1999).

\bibitem{Meade:2008wd}
  P.~Meade, N.~Seiberg and D.~Shih,
  Prog.\ Theor.\ Phys.\ Suppl.\  {\bf 177}, 143 (2009);
  M.~Buican, P.~Meade, N.~Seiberg and D.~Shih,
  JHEP {\bf 0903}, 016 (2009).

\bibitem{Komargodski:2008ax}
  See, for instance, Z.~Komargodski and N.~Seiberg,
  JHEP {\bf 0903}, 072 (2009)
  Z.~Kang, T.~Li, T.~Liu and J.~M.~Yang,
  JHEP {\bf 1204}, 016 (2012).

\bibitem{Ajaib:2012vc}
  P.~Draper, P.~Meade, M.~Reece and D.~Shih,
  Phys.\ Rev.\ D {\bf 85}, 095007 (2012);
  M.~A.~Ajaib, I.~Gogoladze, F.~Nasir and Q.~Shafi,
  Phys.\ Lett.\ B {\bf 713}, 462 (2012).

\bibitem{mh125spart}
  H.~Baer, V.~Barger and A.~Mustafayev,
  Phys.\ Rev.\ D {\bf 85}, 075010 (2012) and
  JHEP {\bf 1205}, 091 (2012).

\bibitem{Babu:2008ge}
  K.~S.~Babu, I.~Gogoladze, M.~U.~Rehman and Q.~Shafi,
  Phys.\ Rev.\ D {\bf 78}, 055017 (2008);
  S.~P.~Martin,
  Phys.\ Rev.\ D {\bf 81}, 035004 (2010);
  P.~W.~Graham, A.~Ismail, S.~Rajendran and P.~Saraswat,
  Phys.\ Rev.\ D {\bf 81}, 055016 (2010).

\bibitem{Martin:2012dg}
  S.~P.~Martin and J.~D.~Wells,
  Phys.\ Rev.\ D {\bf 86}, 035017 (2012).

\bibitem{ISAJET}
  ISAJET program,
  F.~E.~Paige, S.~D.~Protopopescu, H.~Baer and X.~Tata,
  hep-ph/0312045.

\bibitem{Hisano:1992jj}
J.~Hisano, H.~Murayama  , and T.~Yanagida,
  { Nucl. Phys.} {\bf B402}, 46 (1993);
Y.~Yamada,
{ Z. Phys.} {\bf C60} (1993) 83;
 J.~L.~Chkareuli and I.~G.~Gogoladze,
  Phys.\ Rev.\  D {\bf 58}, 055011 (1998).

\bibitem{Pierce:1996zz}
D.~M. Pierce, J.~A. Bagger, K.~T. Matchev, and R.-j. Zhang,
   Nucl.\ Phys. {\bf B491}, 3 (1997).

\bibitem{Moroi}
  T.~Moroi,
  Phys.\ Rev.\ D {\bf 53}, 6565 (1996)
  [Erratum-ibid.\ D {\bf 56}, 4424 (1997)].

\bibitem{gm2}
  G.~W.~Bennett {\it et al.}  [Muon (g-2) Collaboration],
  Phys.\ Rev.\ D {\bf 80}, 052008 (2009).

\bibitem{Baer:2001kn}
  H.~Baer, C.~Balazs, J.~Ferrandis and X.~Tata,
  Phys.\ Rev.\ D {\bf 64}, 035004 (2001).

\bibitem{Hall}
  L.~J.~Hall, R.~Rattazzi and U.~Sarid,
  Phys.\ Rev.\ D {\bf 50}, 7048 (1994);
  R.~Hempfling,
  Phys.\ Rev.\ D {\bf 49}, 6168 (1994);
  M.~S.~Carena, M.~Olechowski, S.~Pokorski and C.~E.~M.~Wagner,
  Nucl.\ Phys.\ B {\bf 426}, 269 (1994).

\bibitem{Gogoladze:2014cha}
  I.~Gogoladze, F.~Nasir, Q.~Shafi and C.~S.~Un,
  Phys.\ Rev.\ D {\bf 90}, 035008 (2014).

\bibitem{Gogoladze:2011db}
  I.~Gogoladze, R.~Khalid, S.~Raza and Q.~Shafi,
  JHEP {\bf 1106}, 117 (2011).

\bibitem{Belanger:2009ti}
  G.~Belanger, F.~Boudjema, A.~Pukhov and R.~K.~Singh,
  JHEP {\bf 0911}, 026 (2009);
H.~Baer, S.~Kraml, S.~Sekmen and H.~Summy,
  JHEP {\bf 0803}, 056 (2008).

\bibitem{pdg}
  K.~A.~Olive {\it et al.}  [Particle Data Group Collaboration],
  Chin.\ Phys.\ C {\bf 38}, 090001 (2014).

\bibitem{isatools}
  H.~Baer, M.~Brhlik, D.~Castano and X.~Tata,
  Phys.\ Rev.\ D {\bf 58}, 015007 (1998);
  J.~K.~Mizukoshi, X.~Tata and Y.~Wang,
  Phys.\ Rev.\ D {\bf 66}, 115003 (2002).


\bibitem{gluinoLHC}
  G.~Aad {\it et al.}  [ATLAS Collaboration],
  JHEP {\bf 1410}, 24 (2014),
  ATLAS-CONF-2013-062 and
  JHEP {\bf 1409}, 103 (2014);
  S.~Chatrchyan {\it et al.}  [CMS Collaboration],
  JHEP {\bf 1210}, 018 (2012) and CMS-PAS-SUS-13-019.

\bibitem{BsMuMu}
  R.~Aaij {\it et al.}  [LHCb Collaboration],
  Phys.\ Rev.\ Lett.\  {\bf 110}, 021801 (2013).

\bibitem{Amhis:2012bh}
  Y.~Amhis {\it et al.}  [Heavy Flavor Averaging Group Collaboration],
  arXiv:1207.1158 [hep-ex].

\bibitem{Asner:2010qj}
  D.~Asner {\it et al.}  [Heavy Flavor Averaging Group Collaboration],
  arXiv:1010.1589 [hep-ex].

\bibitem{Komatsu:2008hk}
  G.~Hinshaw {\it et al.}  [WMAP Collaboration],
  Astrophys.\ J.\ Suppl.\  {\bf 208}, 19 (2013).

\bibitem{Feng:2010ij}
  M.~Viel, J.~Lesgourgues, M.~G.~Haehnelt, S.~Matarrese and A.~Riotto,
  Phys.\ Rev.\ D {\bf 71}, 063534 (2005);
  J.~L.~Feng, M.~Kamionkowski and S.~K.~Lee,
  Phys.\ Rev.\ D {\bf 82}, 015012 (2010).

\bibitem{rges}
  S.~P.~Martin and M.~T.~Vaughn,
  Phys.\ Lett.\ B {\bf 318}, 331 (1993).

\end{thebibliography}
\end{document}